\documentclass[graybox]{svmult}

\usepackage{mathptmx}       
\usepackage{helvet}         
\usepackage{courier}        
\usepackage{type1cm}        
\usepackage{makeidx}         
\usepackage{graphicx}        
\usepackage{multicol}        
\usepackage[bottom]{footmisc}
\usepackage{sidecap}

\makeindex             

\begin{document}
\title*{Logistic Modeling  of a Religious Sect Features }
\author{Marcel   Ausloos}
\institute{Marcel   Ausloos \at Beauvallon, r. Belle Jardini\`ere 483, 
B-4031 Li\`ege, Euroland ,\\ \email{marcel.ausloos@ulg.ac.be}
}
%
%
\maketitle

\abstract*{
 The  financial characteristics of  sects are challenging topics. The present  paper concerns the Antoinist Cult community (ACC),  which has appeared at the end of the 19-th century in Belgium, have had quite an expansion,  and is now decaying. The historical perspective is described in an Appendix. Although surely of marginal importance in religious history, the numerical and analytic description of the  ACC  growth AND decay evolution  {\it per se}  should hopefully  permit generalizations toward behaviors of other sects, with either longer life time, i.e. so called religions or churches, or to others with shorter life time.  Due to the specific aims and rules of the community, in particular the lack of proselytism, and strict acceptance of only anonymous financial gifts, an indirect measure of their member number evolution can only be studied. This is done here first through  the time dependence of new temple inaugurations, between 1910 and 1940. Besides, the  community yearly financial reports can be analyzed. They are legally known between 1920 and 2000.
\noindent
Interestingly, several regimes are seen, with different time spans. The agent based model chosen to describe both temple number and finance evolutions is the Verhulst logistic function taking into account the limited resources of the population. Such a function remarkably fits the number of temple evolution, taking into account a  no construction time gap, historically explained. The empirical Gompertz law  can also be used for fitting  this number of temple evolution data, as shown in an Appendix.  It is thereby concluded that  strong social forces have been acting both in the growth and decay phases.  }
 
\abstract{ 
The  financial characteristics of  sects are challenging topics. The present  paper concerns the Antoinist Cult community (ACC),  which has appeared at the end of the 19-th century in Belgium, have had quite an expansion,  and is now decaying. The historical perspective is described in an Appendix. Although surely of marginal importance in religious history, the numerical and analytic description of the  ACC growth AND decay evolution  {\it per se}   should hopefully  permit generalizations toward behaviors of other sects, with either longer life time, i.e. so called religions or churches, or to others with shorter life time.  Due to the specific aims and rules of the community, in particular the lack of proselytism, and strict acceptance of only anonymous financial gifts, an indirect measure of their member number evolution can only be studied. This is done here first  through  the time dependence of new temple inaugurations, between 1910 and 1940. Besides, the  community yearly financial reports can be analyzed. They are legally known between 1920 and 2000.
 \newline\indent
Interestingly, several regimes are seen, with different time spans. The agent based model chosen to describe both temple number and finance evolutions is the Verhulst logistic function taking into account the limited resources of the population. Such a function remarkably fits the number of temple evolution, taking into account a  no construction time gap, historically explained. The empirical Gompertz law  can also be used for fitting  this number of temple evolution data, as shown in an Appendix.  It is thereby concluded that  strong social forces have been acting both in the growth and decay phases.}

\section{Introduction  }\label{sec:intro}


In recent years, some description of various communities, in particular by physicists \cite{stauffer04,carbone,chakr06,mimkes,galam08},  has been of interest  quite outside sociology\footnote{At this level, one should 
pay some respect to earlier work, i.e.  in 1974, Montroll and Badger \cite{MontrollBadger74}  have  introduced, into social phenomena,  quantitative approaches,  as physicists should do along mechanics lines.  Indeed communities can be considered as in a thermodynamic state,  - as Boltzmann discovered \cite{Blackmore 1995} after Comte \cite{Congrev 2009}, 
},
  since the communities  are made of $agents$,  defined by  several {\it degrees of freedom}, like sex, age, race, citizenship, wealth, intellectual quotient, love for music groups, sport activity, language, religion, etc.   

Interest in religious movements  has much led  scholars  to consider  those through field observations  rather than  through surveys. Studies on 
Bruderhof \cite{Zablocki 1971}, Jehovah's Witnesses \cite{Beckford 1975}, 
Satanist groups \cite{Bainbridge 1978}, the Unification Church \cite{Shupe and Bromley 1979},  Scientology \cite{Wallis 1977},  the Divine Light Mission \cite{Downton 1979}, and other movements \cite{Glock and Bellah 1976}  have focussed more on  the importance of symbolism, ritual, and discourse in the construction of "religious meanings" \cite{Bird 1979,Tipton 1982,Westley 1978} than on the motivation within  social, financial or general sectary aspects.  

Of course, Marx and Engels  \cite{MxEng} did ask   to what extent   religion serves as an opiate that stifles social change, but this political view emphasizes exogenous goals and means.  Durkheim  \cite{durkheim} 
did explore the ''degree'' to which  a cult, later on a religion, through a church,  entices a source of cohesion that stimulates  agent collective endogenous actions. One should claim that  such a ''degree'' contains some measure of a (at least, moral) satisfaction.  Still, too much consideration on the various goals of  religious organizations, i.e. saving the body OR (my emphasis) the soul,  
 might  obliterate an objective/quantitative  approach.  Thus, not disregarding the need for  a qualitative understanding of  the complex role of religions in social service delivery performance,
  a more quantitative approach, along agent based modeling considerations,  might  shine some light on  successes and failures of some cult or sect.

In fact, within Comte ideas \cite{Congrev 2009}, one has  often  attempted to describe economic  and sociological features within some analytic equations, involving agents,  (i) with degrees of freedom, thus interacting with ''external fields'', like a spin with a magnetic field or a charge with an electric field, $and$ (ii) interacting with each other within some cluster. External fields  can be mapped into so called  {\it social forces}   \cite{PNAS75.78.4633-7-Montroll-socialforces} through a change of vocabulary. For simulating the numerical evolution in size of a religious movement,  many available models of opinion dynamics are also available,  several taking into account preferential attachment, - seen in  \cite{religion1_505,religion2_555,religion3_566,religion4_568} as one of the fundamental dynamical causes of the evolution of such religious movements.  

Beside such considerations,  one may  wonder  about financial and/or economic aspects of   religious communities, and how they evolve in some so called market.
There is a huge literature, going back to  \cite{ianna98}, and much intense work on the economics of religious adepts and their hierarchy; both at the micro- and macro-levels. Many interesting considerations exist and are worth to be read, but  quantitative modeling  is somewhat absent, 
 as well as some  search for empirical laws, and for explaining them,  say mathematically.

In view of  the above, I was interested in finding whether one could get some economic or financial data on the evolution of a community made of agents having a well defined  so called ''degree of freedom'', like their religious adhesion. For various reasons, this is not so easy to find:  such  communities do not want to be appreciated as rich, or  as poor.  Financial data are rarely released. 
 There are  known  psychological difficulties in  merging  considerations about  money and religion  \cite{ianna98}.  One may wonder why!

One crucial request, for a study within the so defined framework, is to find some community for  which   growth  and decay have been observed, but have not been influenced by too many, violent or not,    competitive aspects with other cults. Therefore,  the time life of an interesting community should be rather short, yet long enough to have some meaningful data.

A community like the Antoinists, here below called  the Antoinist Cult Community (ACC),  exists for about more than a century, has markedly grown,  \cite{Debouxhtaybook,Dericquebourg} and is now apparently decaying. For the reader information, some comment on the community origin and roles is left for  Appendix A.  In France, the  religious association community is considered as a sect  \cite{JOFrance} 
but it is a {\it Etablissement d'utilit\'e publique} (Organization of Public Utility)\footnote{Such a legal association status seems to have been invented for the Antoinist Cult in Belgium, though   no formal proof of the latter can be found in notes of the parliament related to the  1921 law elaboration process.    Notice that the Minister in charge of the application is the Minister of Justice, - apparently due to suspicion going on with such unfamiliar religious/charity matters.   }
 in Belgium, since 1922.  For short, the ACC will be called a religious sect, nevertheless.   

 It would have been of interest to have some quantitative measure of the true number of adepts, in order to relate such a data with previous studies, as in  \cite{religion1_505,religion2_555,religion3_566,religion4_568}. However, one  remarkable and quite respectable characteristics of the ACC, beside $not$ to have any proselytism action, unlike the more financially active religious movements, is not to keep  anything, like a financial gift,  which would induce some private information.  Thus,  information on adept number evolution can only come from indirect measures.   The presently relevant, thereafter studied,  data  is limited to: (i) Sect.  \ref{sec:templedataset},   the number of temples, built in Belgium,  before World War II, and  (ii) Sect.  \ref{sec:financialdataset},  the financial activity of the community, i.e. yearly income and expenses, for over about 80 years. In Sect. \ref{VLLasABM},  the most commonly accepted  kinetic  growth law  for describing a population evolution, i.e. Verhulst law  \cite{Verhulst845}, is introduced as the analytic solution of a possibly relevant agent based model (ABM).

 Interestingly, the growth in the  number of temples  will be  found to follow   such a Verhulst law  \cite{Verhulst845}, in Sect. \ref{ANTEVLL}.  In estimating acceleration and deceleration processes in the inauguration of temples, one will observe and quantify some social force effect, in the sense of Montroll \cite{PNAS75.78.4633-7-Montroll-socialforces}, in Sect. \ref{summary}.  Some ABM  interpretation   is given, in Sect. \ref{ABMI}.

 In Sect. \ref{sec:fits},
the  evolution of  financial  data is  
  also studied   starting from fits 
  according to  the Verhulst logistic function  \cite{Verhulst845}. 
 A  succession of three logistic regimes is found, in Sect. \ref{sec:fitsNAIncexp}, both  for  income and expenses  data.

 Since there is some  natural interest in considering both growth and  decay processes of  a community, 
  the decay  of income and expenses after 1980  is also examined, in Sect. \ref{decay}. 
 A  trivially  simple behavior is  $not$ found though. 
The non symmetric time dependence of growth and decay regimes leads to an open question.

 Sect. \ref {sec:conclusions} serves 
as a conclusion:  the complexity of quantitatively studying a religious community through its social history   will be  emphasized. Indeed several growth regimes are found. 
 On the other hand, the recent  decay process is hardly mapped into a simple analytical form. 
 This is in marked contrast with ecological or laboratory based data on population evolutions. Thus, {\it as should most likely be  really expected}, it is concluded that social phenomena are very complex processes offering much challenge  for  future physics modelling,  suggesting challenging investigations through agent based model simulations. 

Note that the Verhulst mapping is sometimes criticized as an unrealistic, too simple, model. Therefore,  
  the sometimes considered as the best alternative  Gompertz (human mortality) law   \cite{Gompertz825} is adapted to the present context, in Appendix B, -   tuning the parameters into their size growth rather than decay value. 
     $Only$    the number of temples evolution case is  
    reported within such a Gompertz approach; see App. B. 
Finally, in  App. C, it is emphasized that social forces can be introduced at least in two different ways in an ABM,  based on Vehulst and/or Gompertz analytic evolutions.

 As hinted, this paper is based on a compilation of  already publishedÊ papers on  
econophysics aspects of the Belgium ACC \cite{577PhA391.12.3190Ð97auslantoin,579,582}. but  new figures are here included.

\section{The Data Set} 
 \label{sec:dataset}

\subsection{   Number of Temples Data}  \label{sec:templedataset}

Since there is no data on adept adhesions, one indirect way of observing the evolution of the size of the ACC has been mapped into  the counting of the number of temples as a function of time.  Such data is meaningfully available for Belgium  till 1935.
   Note that temples have been mainly constructed in Belgium, though others exist in France and Brazil.  Though it might also be of interest to consider the data for the whole sect on a world wide basis, only the 27 temples constructed in Belgium during the main growth phase of the  cult  are here below considered, for coherence.  
The number of temples constructed as a function of time has been extracted  both from the archives of the ACC and from \cite{Debouxhtaybook} and \cite{Dericquebourg} compilation and discussions.

   Most of the times, the exact day of the $inauguration$  or $consecration$ is known, - sometimes  (twice) only the month is known. To be more precise about  the exact day of  the event, for the latter cases, would request much time consuming, searching for this information through news media, without being certain of the success. When in doubt, the dates in an Appendix  of a 1934 book by Debouxhtay \cite{Debouxhtaybook} are used,  because I consider them as the most reliable ones. In order to count the number of days and months between successive rise of temples, when the day  is unknown, the date has been assumed to be   the 15th of the month. The number of days  between two events has been calculated, taking into account bissextile years if necessary; the number of months has also been calculated but rounded up to the nearest integer according to the sum between the number of days till the end of the first month and from the beginning of the last month corresponding to the two marginal events so considered. The $x$-axis for the figure discussed will thus   be the cumulative number of months since the rise of the first temple  in Jemeppe-sur-Meuse, on Aug. 15, 1910; the data extends up to the 27-th temple raised\footnote{for 135 880.70 BEF of that year   }  in  La Louvi\`ere, on Dec. 03, 1933.  After a huge consecration gap, lasting more than 20 years, four other temples have been raised in more recent years,  but three of these have been closed already. These temples are not included in the analysis.
   
    For  giving some perspective, let it be noticed that, in France,  the first temple  was consecrated  in Paris (13\`eme) on Oct. 26, 1913, while the 15th was in  Valenciennes, on Aug. 07, 1932. The most recent one, the 31st and  32nd, were inaugurated in  Caen and Toulouse in  1991 and 1993 respectively, but, even after some local research, the days and months are unknown. Thus, the evolution of the number of temples in France is not studied here because of  such uncertainties.

\subsection{ Financial Data}    \label{sec:financialdataset}
\begin{figure}
\centering
  \includegraphics[height=5.8cm,width=5.8cm]{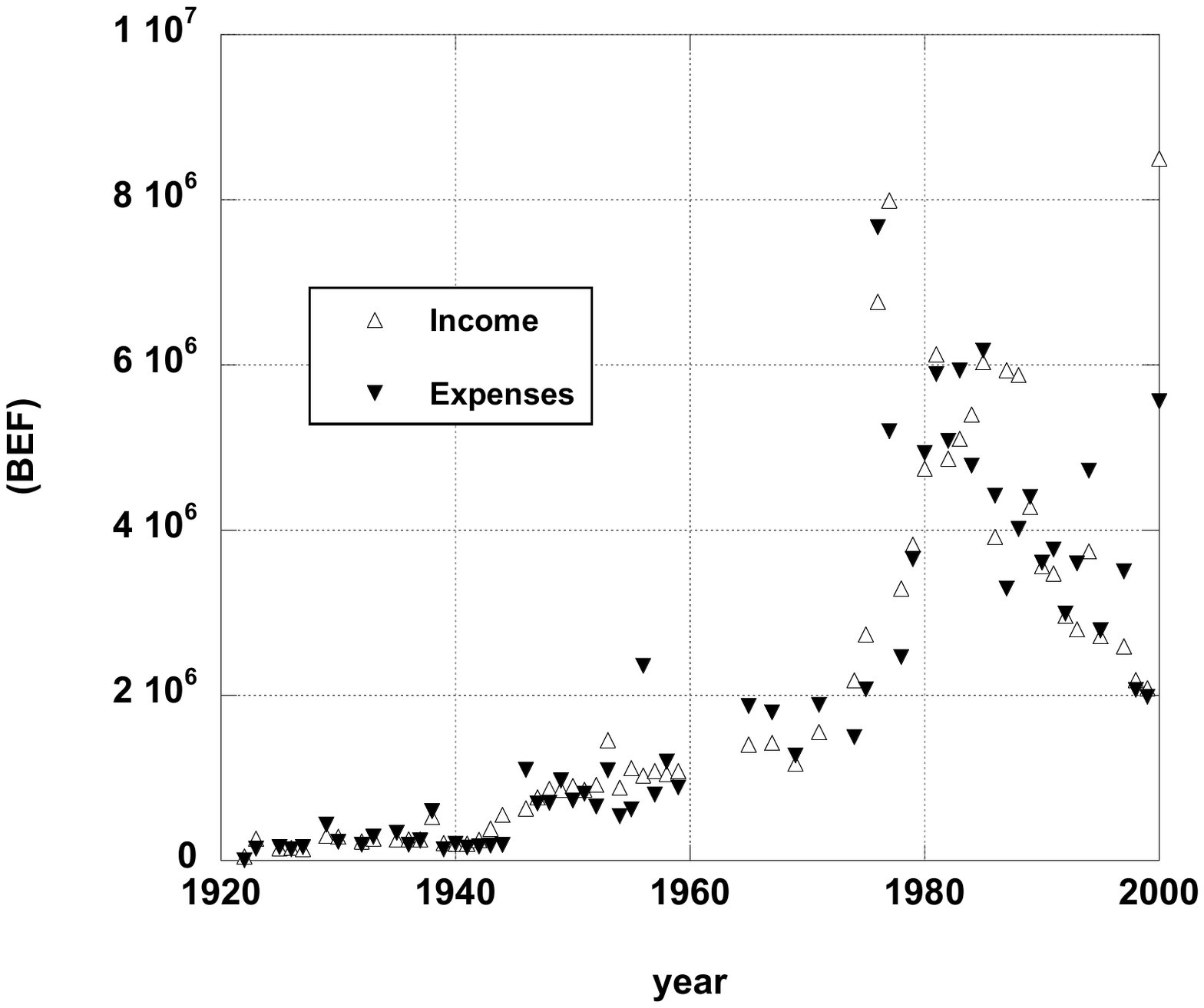}
 \includegraphics[height=5.8cm,width=5.8cm]{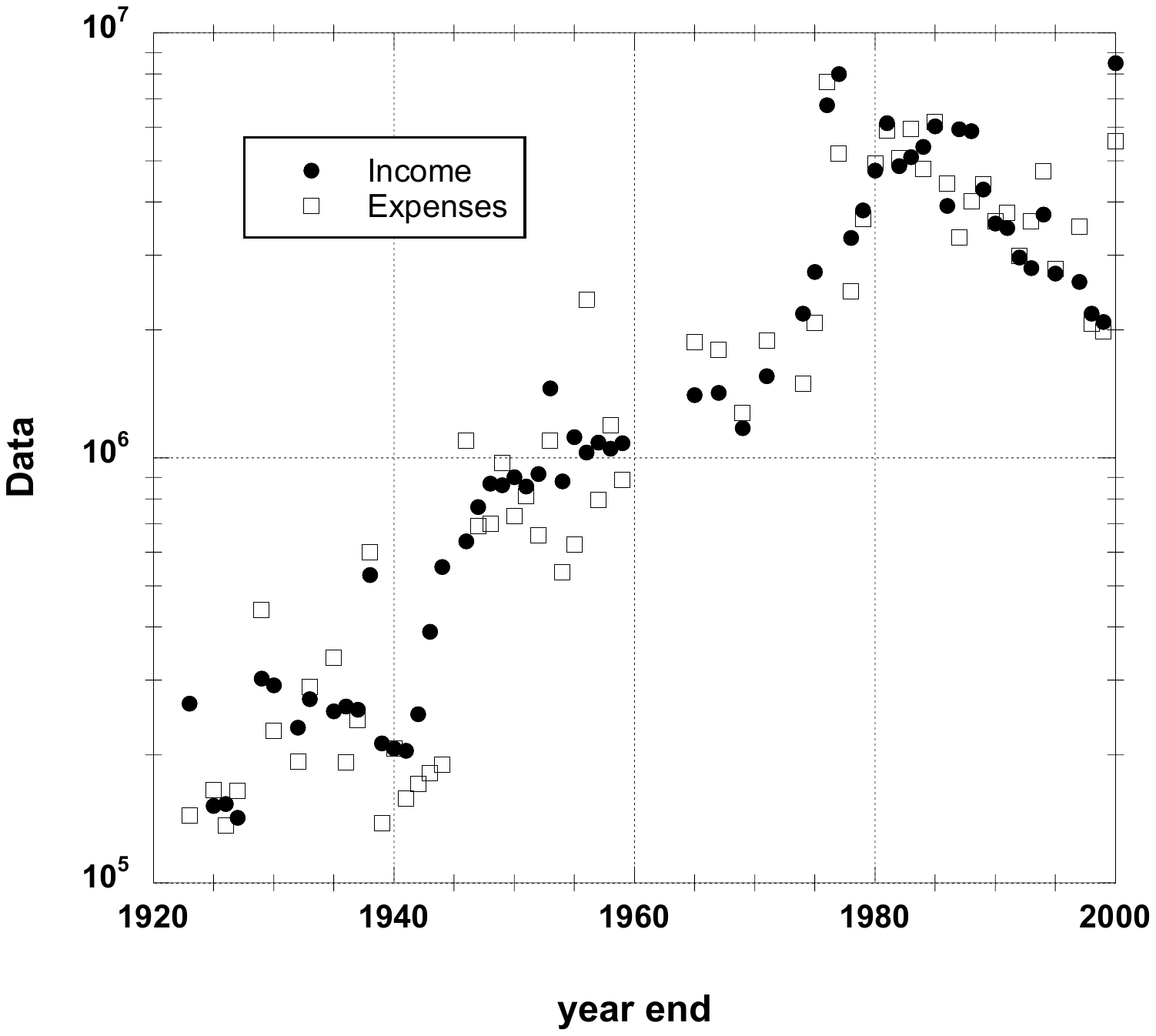}
\caption{   (left) Yearly income and expenses of the Belgium  Antoinist Cult Community  as reported in the {\it Moniteur Belge}; (right)  logarithmic scale display  of yearly income and expenses  suggesting three growth regimes before  some marked decay  }
\label{fig:incomexpenses}
\end{figure}

The financial data set  has been extracted from the  Belgian daily official journal, i.e. {\it Moniteur Belge}, when it was available in the archives of the Antoinist Cult Library in Jemeppe-sur-Meuse. A few  issues  are missing, i.e. $ca. $ 1960-1965, without any known reason,  but those  do not appear {\it a posteriori}, from the subsequent data  analysis, to impair the discussion and conclusion.
  The examined time range  starts in 1920, 
   i.e. since when it was mandatory to report it. However, due to the introduction of the EUR in 2000, in order to keep the Belgian Franc (BEF)  as the usual unit, only the data till 2000 is  discussed, - again without much apparent loss  of content for the present discussion and conclusion.
   Due to an evolution in the Belgian legal rules for reporting income and expenses data over the last century,  many detailed items can be found in the oldest reports.   In  order to have enough meaningful comparison over several decades, such  detailed data on the years, i.e. grossly before 1940, has been concatenated such that only the yearly  total  $incomes$ and total $expenses$ are finally used,  are displayed  in Fig.\ref {fig:incomexpenses}, ???? and discussed  below. Notice that  these  so called  $income$ values do not take into account the left-over from the pervious year(s).  That is why, in  Fig.\ref {fig:incomexpenses}, sometimes, it seems that there are more expenses    than income in a given year.
   
    Moreover, the   financial  data, - equivalent to the Belgium community  data, here studied, but corresponding to the France Antoinist cult activity, is not available, and thus cannot be studied below.  
   
   \subsection{ Agent Based Model Numerical Analysis Methodology. The Verhulst Logistic Function }\label{VLLasABM}

  For the modelization of an agent based community growth as a function of time $t$,   let us take the (Verhulst) so called logistic  function as the first approximation, i.e.,  a sigmoid curve, 
\begin{equation} \label {Verhulst2sol}
z= z_{\infty} \frac{e^{r(t-t_m)}}{1+e^{r(t-t_m)}}\; =  \frac{z_{\infty}}{1+e^{-r(t-t_m)}}\;  ,
\end{equation}
 where $z_{\infty}$  is the upper limit  of $z$  as time $t$ tends to infinity, $t_m$ is  the position  of the inflection point, at   mid   $z$ amplitude, such that $z_m=z_{\infty}/2$, and $r$ is the supposedly constant growth rate.  This way of expressing the logistic curve has the advantage that the initial measure $z_0$,   at $t=t_0$, is a rapidly fixed value:
  $z_0=z_{\infty} \frac{e^{r(t-t_m)}}{1+e^{r(t_0-t_m)}}\; =  \frac{2¥.z_m}{1+e^{-r(t_0-t_m)}}\; $, for estimating one of the three parameters in Eq. (\ref {Verhulst2sol}).
 
 Above the inflexion point,  one can use an asymptotic expansion i.e.
\begin{equation} \label {Verhulst2solup}
z \simeq z_{\infty}(1-e^{-r(t-t_m)})\;  .
\end{equation}
 Such a latter exponential growth behavior, Eq.(\ref{Verhulst2solup}),  
  is sometimes referred to as the von Bertalanffy curve  \cite{vonBertalanffy} curve, -  of mass accumulation, in biology. It is also through an appropriate change of variable, nothing else than Malthus exponential growth rate,  $y=y_0\;e^{b\tau}$, i.e. if  $r\equiv -b$, $\tau \equivÊ t-t_m$ and $z\equiv   z_{\infty} (1 -y/y_0)$.
  
  Note also that 
  \begin{equation} \label {VerhulstMontrollform}
\frac{z/z_{\infty}}{Ê1-  z/z_{\infty}}=  e^{r(t-t_m)}\;  .
\end{equation}

\section{  Number of Temples Evolution }\label{ANTE}

 The number of raised temples by the Belgium ACC is displayed in Fig. \ref{fig:PlotV30ALMFtemplefitsbw}, as a function  of the evolved month since the rise of the first  temple on Aug. 15, 1910.    For the sake of completeness, let it be mentioned that Dericquebourg \cite{Dericquebourg} has given a sketch of this number of temple evolution, in Belgium and in France, in an appendix to his book, though without any  qualitative nor quantitative discussion.
 
 \subsection{ Numerical Analysis. Verhulst Logistic Law } \label{ANTEVLL}
 
\begin{SCfigure}[\sidecaptionrelwidth][h!]
\centering
\caption{Logistic fits,  at low and high $t$,  of the number of temples of the Antoinist Cult Community in Belgium as a function of the number of months,  since the consecration of the first temple on Aug. 15, 1910}
  \includegraphics[height=6.65cm,width=8.6cm]{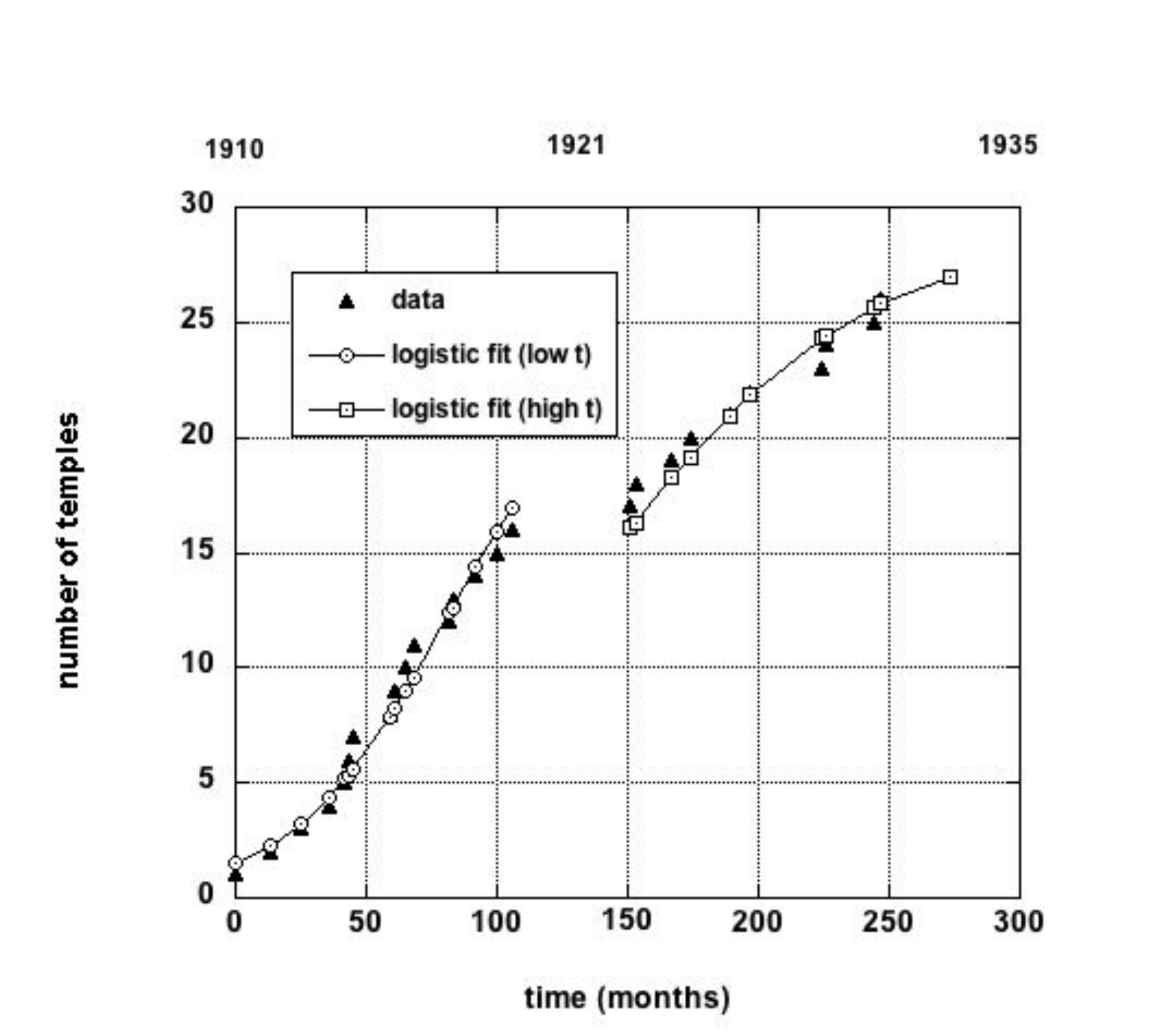}
\label{fig:PlotV30ALMFtemplefitsbw}
\end{SCfigure}

 The best fit  behavior  to a logistic law has been searched through a log-log plot method, based on Eq. (\ref {VerhulstMontrollform}). The upper value, $z_{\infty}$,  was imposed to be an integer.  It has occurred after many simulations, but it could occur readily to many,  that $two$ distinct regimes must be considered: one at {\it ''low t''}  , i.e. during the initial growth of the ACC, and another in   {\it high t}, at ''later times''. The regimes are readily separated by a 4 year gap, between 1919 and 1923, during which there was $no$ temple construction.
The low $t$  logistic fit of the number  ($c$) of temples as a function of the number ($m$) of months (cumulated since the rise of the first temple)  
corresponds to
\begin{equation}
\label{Vlowt}
   c(m) = \frac{24}{1+e^{-0.03395*(m-80)}}
\end{equation}
 while the fit in the upper $m$ regime  corresponds to
 \begin{equation}
\label{Vhight}
  c(m) = \frac{29}{1+e^{-0.0195*(m-140)}}.
\end{equation} 
Both data and fits, in  each    regime, are combined and shown  in  Fig. \ref{fig:PlotV30ALMFtemplefitsbw}.

 It is remarkable that the initial growth rate for such data is about  0.03395, i.e. largely more than 3 temples every ten years, and reduces to 0.0195 in the latest years, i.e. about 2 temples per year. Nevertheless, although the initial logistic law should have led to expecting $\sim$ 24 temples at saturation, the latter one would predict 29 temples at most. It is interesting to recall here that Verhulst modification of Malthus (unlimited growth) equation was based on considering a ''limiting carrying capacity'' of the "country",  for the considered population. {\it Mutatis mutandis}, such 24 and 29 $z_{\infty}$ values reflect  such an effect.
 
 Notice that a  $unique$ logistic curve fit,  over the whole time interval, would give a value of the growth rate $\sim$ 0.02355, but not fulfilling the Jarque-Bera (JB) test    \cite{jbtest}, - even when  finite sample effects are taken into account   \cite{ebeling280}.

 One might debate whether the  original logistic map is the most appropriate law to be considered. One might  suggest a skewed logistic with extra parameters, as considered, e.g.  in    \cite{ZwanzigPNAS70,Pearletal28f,Pearletal28a} 
  studying  various population growth cases. This has not been considered for this report, because the parameters entering such skewed curves are hardly meaningfully interpreted, in the present  investigation context. 
  It seems preferable in view of the data analysis and the framework of this investigation to further discuss the findings, in  Sect.  \ref{summary}, as  due to the influence of social forces  \cite{PNAS75.78.4633-7-Montroll-socialforces} on agents, i.e. adepts. 

Note also that the Gompertz (double exponential) growth law  \cite{Gompertz825} is studied in App. B, for the above data.

\subsection{ Quantitative Measure of Social Forces}\label{summary}

At first sight, the presently investigated growth regimes do not seem  to overlap much. Moreover the rates of growth seem somewhat different in the successive regimes, indicating  sequential rather than overlapping (and competitive)  processes, in contrast to social and technological cases \cite{RS13.07.5fokas,Marchetti1997b}, as well as botanical  \cite{JHSBbilogist,PNAS6.20.397Reed} and other biological, - in which successive competing molecular reactions are of course involved .
 Therefore, one might rather consider, beside Òsocial endogenous contagionÓ,  an effect due to, in the words of  Montroll, "social forces"  \cite{PNAS75.78.4633-7-Montroll-socialforces}. It is worth to recall that Montroll argued that social  evolutionary processes occur due to competition between new ideas and old ones.   Moreover, deviations  from the classical logistic map are often associated with  intermittent events. In many cases, a few years after on such event, it can be abstracted
as an instantaneous  function impulse.  Montroll   argued that the most simple generalization of Verhulst equation,  in such a respect, goes when introducing a force impulse,  $F(t) = \alpha\;\delta(t-\tau) $, in the r.h.s.  of  the kinetic  Verhulst equation    so that the dynamical equation for some forced evolution process $X (\equiv z/z_{\infty})$ reads 
\begin{equation}
\label{Montrolleq}
\frac {d\;ln(X)}{dt }= k(1- X) +   \alpha\;\delta(t-\tau) .
\end{equation}

\begin{SCfigure}[\sidecaptionrelwidth][h!] 
\centering
  \includegraphics[height=6.65cm,width=8.6cm]{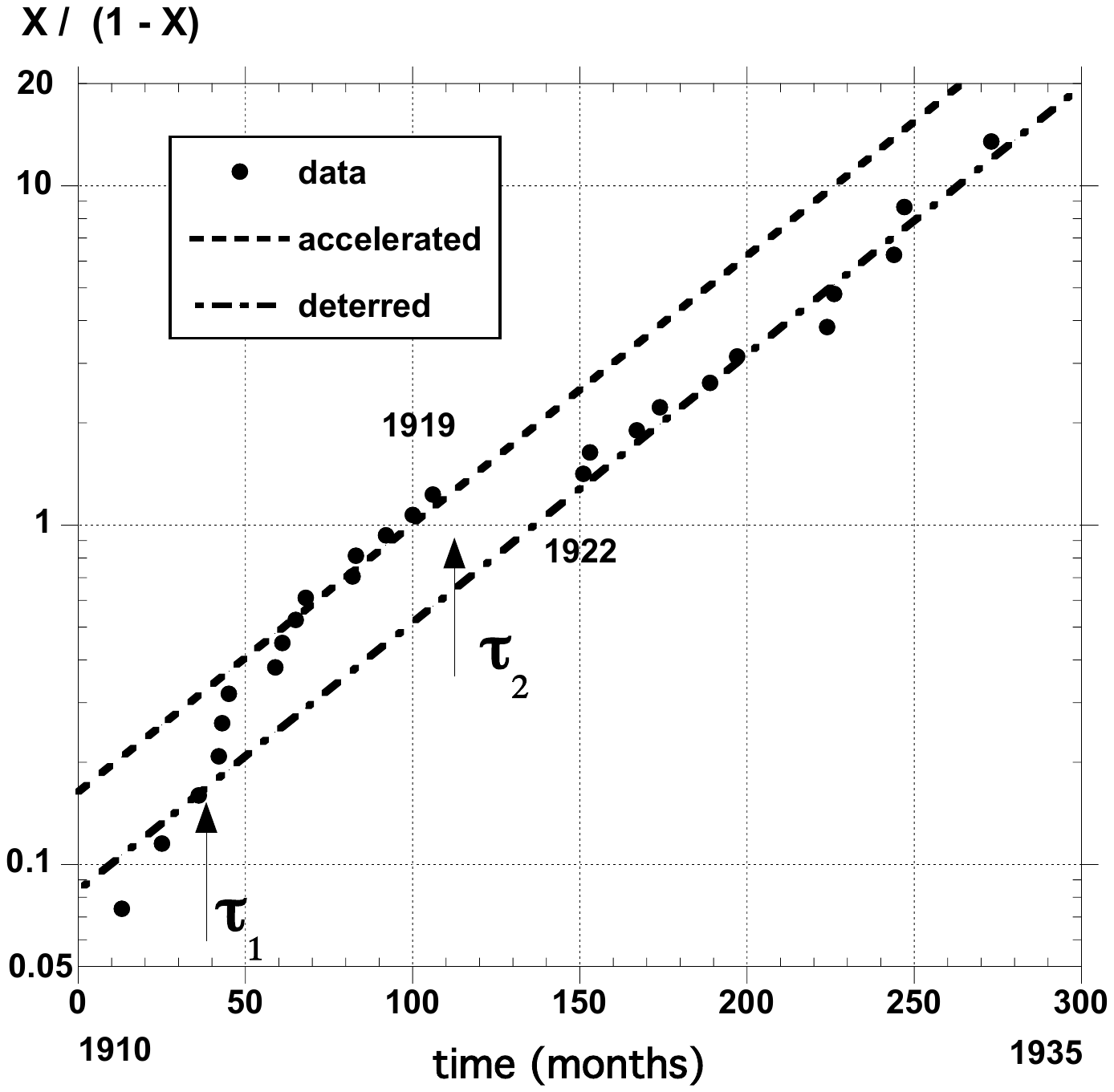}
\caption{  Logistic variation    ($X/(1-X)$) of the number of  ACC temples ($X$)  in Belgium as a function of the number of months (cumulated from the raise of the first temple, in 1910), indicating   a ""social force effect" at time $\tau_1$, accelerating the process over a time span, 
and a decelerating force impulse at time  $\tau_2$
 }
\label{fig:Plot53templeforceAB}
\end{SCfigure}

In so doing, in the time regime after the withdrawal of  the intermittent
force, the   evolutionary curve are parallel lines, on a semi-log plot, see Eq.(\ref {VerhulstMontrollform}):  the unaccelerated one, above or below the latter depending whether the process is accelerated or deterred at time $\tau$.    The impulse parameter $\alpha$  is easily obtained as explained in  \cite{PNAS75.78.4633-7-Montroll-socialforces}  and  for the present case in  \cite{582}. One finds $\alpha_1= 0.103$, at $\tau_1\sim 1914$,  and $\alpha_2=  - 0.405$, at $\tau_2\sim 1922$, from Fig. \ref{fig:Plot53templeforceAB}. These are very reasonable orders of magnitude. It should have been obvious that the decelerating force should be higher in magnitude that the 	accelerating one.   The fact that the forces are usually not instantaneous ones, and do not    suddenly accelerate or decelerate the process, are approximations which are   reexamined, in App. C, - where some emphasis is further made on different points of view: $mathematical-like$, at first sight, but fundamentally relevant for  discussing the causes of evolution of many populations, along ABM ideas. 

\subsection{Agent Based Model Interpretation} \label{ABMI}
The  ACC  present  hierarchy interpretation  and mine go along the lines of the historical points reported in the Introduction here above. The first acceleration, ca. 1914  can be historically connected to the first world war. The workers and their families needed some intra-community social support, and interestingly being satisfied by the   healing of their soul and sometimes body, gave quite an amount of money to build structures, temples,  replacing the mere ''lecture rooms'' where the adepts first gathered for   cult activities.  

After the war,  income and housing taxes  were implemented.  However, social organizations attempted to be legally screened from such taxes. In Belgium, since its independence from The Netherlands, in 1830, the catholic priests, and  officials of  a few other cults nowadays, are paid as government employees, on a specially adapted scale.  However, the introduction, and legal recognition, of  a new active socio-religious group into the  religious affairs of the country was not well appreciated by the catholic  church leaders and adepts.  Whence it took a while before the Belgian government, manipulated by the bishops and  catholic members of the parliament, accepted to consider a new law  establishing a role for social organizations, associated to some religious movement. During more than four years, as convincingly outlined  by Dericquebourg   \cite{Dericquebourg}, the intended law\footnote{Nowadays,  the law is applied much outside socio-religious groups, e.g. to museums.}  on Organisations of Public Utility suffered many parliamentary delays, starting from 1919,  {\it in fine} much decelerating the temple construction process. One should admit that it was quite natural  from a tax point of view  to wait for the rising of new temples. When the law was finally voted and applicable, accumulated money could be used for temple construction, over a few months (20 or so), after 1923.  The growth could resume, as seen in Fig. \ref{fig:PlotV30ALMFtemplefitsbw}. 
 For completeness, note that the 1929 October crash occurs during the 230-th month; observe the small gap around such a time.

Later on, after 1938, the number of temples did not much increase,  indirectly indicating the   unnecessary need for  such constructions because of the stability or even decay in the number of adepts. The  recent closing of three temples  seems to confirm such a number evolution.

Following such considerations,  the relationship between psychological and sociological needs, at times of great economic difficulty,  can be observed in such a cult adepts.  It is remarkable that poor social conditions led to high cash gifts, - sufficiently as to build temples during a war time.  The somewhat incredible  rising level of gifts by adepts for the construction of temples at difficult times is an interesting observation of the intracommunity  {\it autocatalytic process}\footnote{The velocity of growth, or concentration growth, of an enzyme is depending on the concentration of substrate \cite{Menten and Michaelis 1913}. the rate of an enzyme-catalyzed reaction is proportional to the concentration of the enzymeÐsubstrate complex}. Yet, it  could be argued that the ACC  hierarchy  was intending to build more temples in order to increase the acceptance of adepts, as  ''clients'', as done in more financially prone churches nowadays.  However, it should be re-emphasized that proselytism  is far from any  such goal  in the ACC. Let it be repeated that the cult ''desservants"" are not paid. The (lack of) growth in the temple number is thus entirely due to the intra-community factor state, rather than a leadership manipulation.

\section{  Income and Expenses Evolution } \label{sec:fits}

Recall that yearly  expenses and income over about 80 years during the 20-th century. Note that this so called   "income"  value does not take into account the left-over from the pervious year(s).    Further study has led to interesting  analytic description and explanations about the evolutions of  these financial matters \cite{577PhA391.12.3190Ð97auslantoin}. 

 \begin{figure} 
\centering
\includegraphics[height=6.8cm,width=5.8cm]{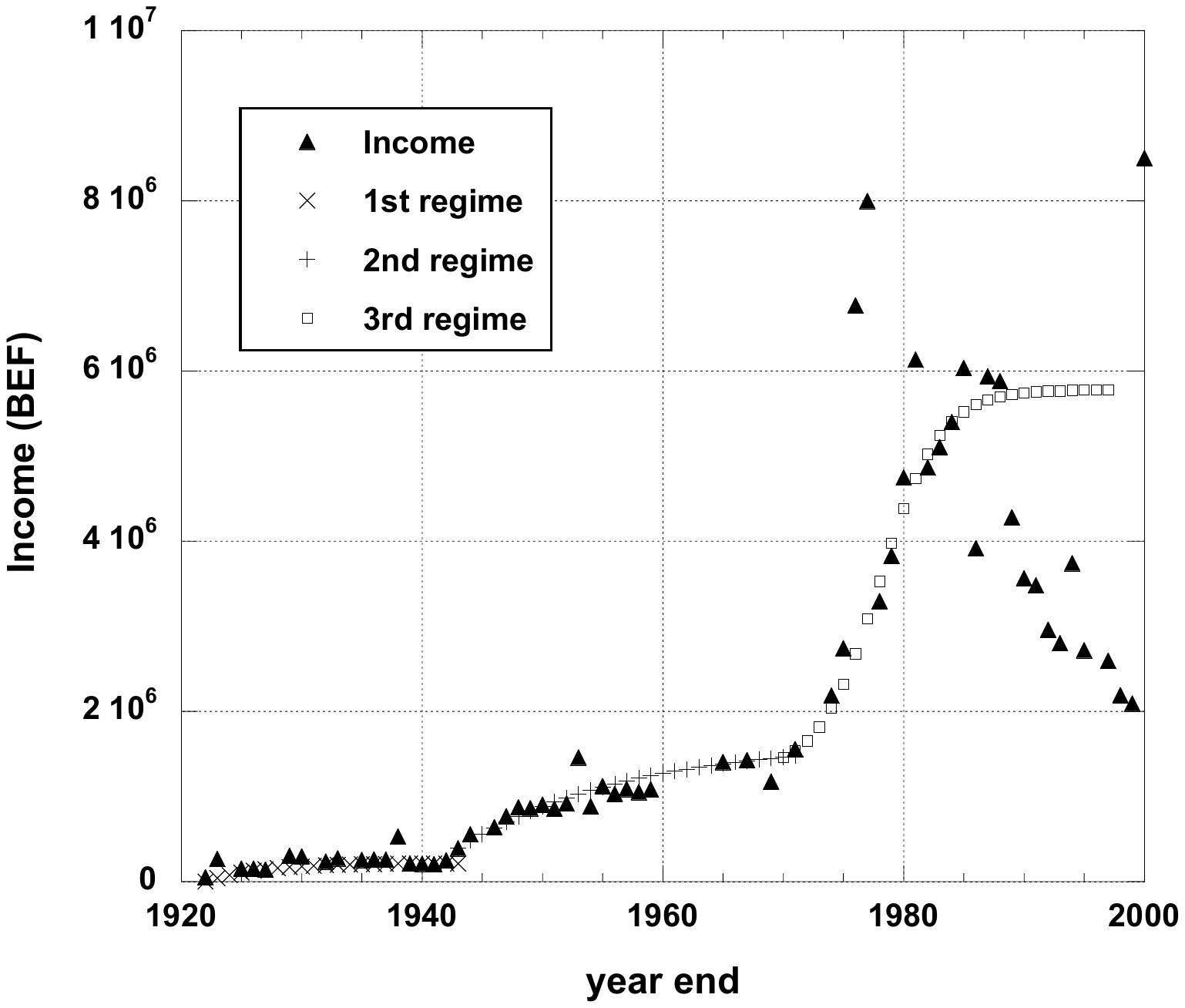}
 \includegraphics[height=6.8cm,width=5.8cm]{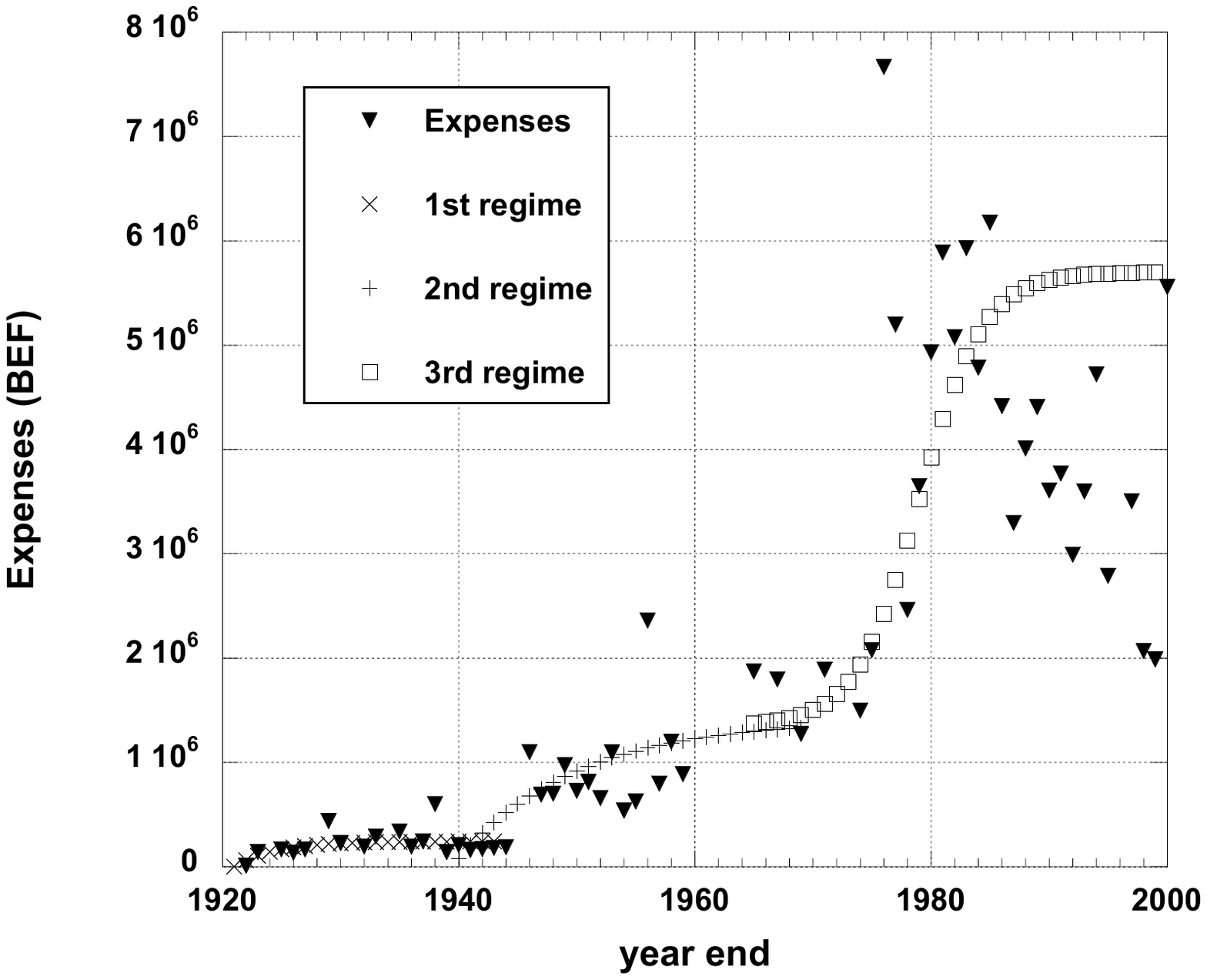}
\caption{Successive logistic fits of the  belgian Antoinism Cult Community  income  (left) and expenses (right) on mentioned year, taken from  official  {\it Moniteur Belge} journal, indicating three time dependent regimes; parameter values are found in Table 1  }
\label{fig:Plot5EFJforC3fitsbw}
\end{figure}

\subsection{  Numerical Analysis.  } \label{sec:fitsNAIncexp}

The raw data, Fig. \ref{fig:incomexpenses}
  appears as if points are pretty scattered. However, after much fit searching, it appears that in $both$ income and expenses cases, three growth and one decay  regimes can be found, see 
  Fig. \ref{fig:Plot5EFJforC3fitsbw}, approximately over barely overlapping time spans. 
Because of the  $y-$axis scale, crushing the first two regimes, it is fair to describe the figures in words.
 The presence of a maximum in 1929 can be  observed before a smooth and short decay till the beginning of the first world war in 1940.  The next   growth regimes ends with a small decay at the end of the {\it golden sixties}. 
Finally,  the last bump  is  seen  to occur   at the beginning of the eighties. A marked decay follows thereafter. 
 
\begin{SCfigure}[\sidecaptionrelwidth][h!]
\centering
  \includegraphics[height=6.5cm,width= 8.5cm]{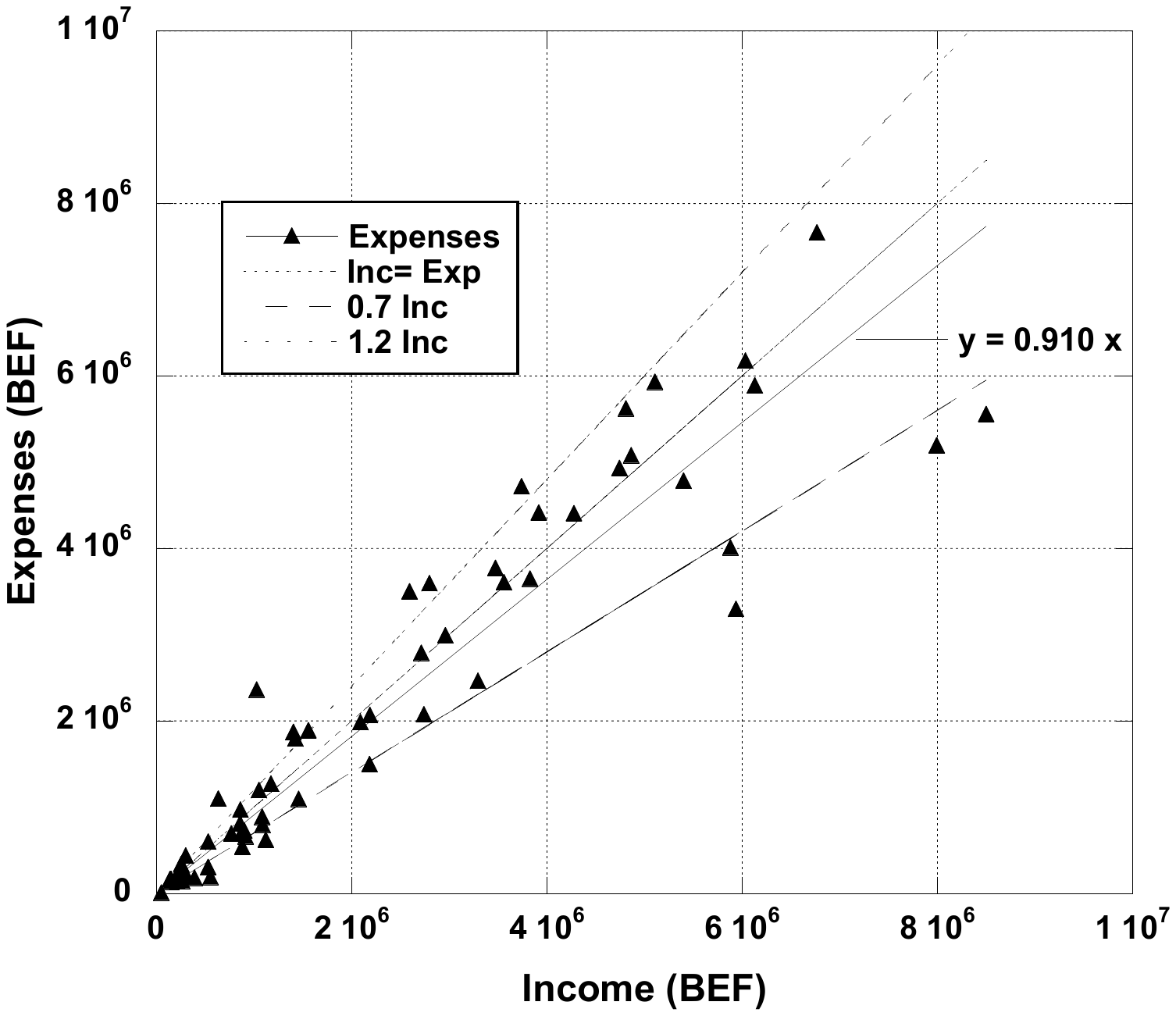}
\caption{Apparent $quasi$-linear correlation between  the   Belgium ACC expenses  and income, demonstrated by: the best linear fit line going through the origin, the 45$^\circ$ slope line, and two linear envelopes  }
\label{fig:Plot6data2CDlinearlawbw}
\end{SCfigure}

  \begin{table}\label{incomexpensfits}\begin{center} \begin{tabular}{|c|c|c|c|c|c|c|c| l |    }
  \hline
\multicolumn{4}{|c|}{income }&&\multicolumn{4}{|c|}{expenses   }\\
\hline
\hline     time  regimes& $ z_{\infty} \; 10^{-6}$ & $r $&	$t_m$&&time regimes  &$ z_{\infty} \; 10^{-6}$ &$r$  &$t_m$   \\
\hline \hline $1922-1940 $ &0.24& 0.29 &  1921   &&   1922-1946&0.22 &0.22&1922 \\
\hline $1940-1968$  &1.20&  0.10 & 1941    &&   1946-1968& 1.45&0.08  &1946\\
\hline $1968-1980  $&4.50&   0.63 &1979    &&1968-1980& 4.50&0.50 &1978\\ 
\hline \end{tabular}  \end{center}

\caption{ Comparison of the growth rate $r$  in yearly income and expenses of the Belgian Antoinist Cult Community in different  time regimes; the growth rates correspond to the respective  regimes in Fig. \ref{fig:Plot5EFJforC3fitsbw};  
$t_m$  and $z_{\infty}$ are fit parameters  for the logistic function, Eq.(\ref{Verhulst2sol}),  see text}
\end{table}
 
The fit parameter values are given in Table 1. 
 In general, it appears that the  income  growth rate  is slightly larger than  the expenses growth rate; the more so except in the 3-rd regime, where the  $absolute$ difference is much larger. However, let it be stressed that  the growth rate difference, in $relative$ value, is close to  27\%, -  in $each$ case.  
 Such correlations between  expenses and  income  can be seen in Fig. \ref{fig:Plot6data2CDlinearlawbw}. The relationship law is approximately linear,  though sometimes, a few large income  years can occur.  
 
 It can be observed that a  one year time delay occurs in such cases,  whence leading to  conclude about sound financial management  by the  ACC hierarchy.

 \subsection{ Most  Recent Decay Regime 
 }\label{decay}

A sharp peak in the income and the expenses data is observed near 1980.\footnote{One should remember that the available budget  for expenses in a given year is the sum of the expected  income for the year and the left-over from the previous year.} The decay  has been searched whether it obeys a simple law.\footnote{Due to the legal change in reporting data after 2000, after the introduction of the EUR, only data previous to 2000 is  considered.} 

\begin{SCfigure}[\sidecaptionrelwidth][h!]
\centering
 \includegraphics[height=6.65cm,width=8.6cm]{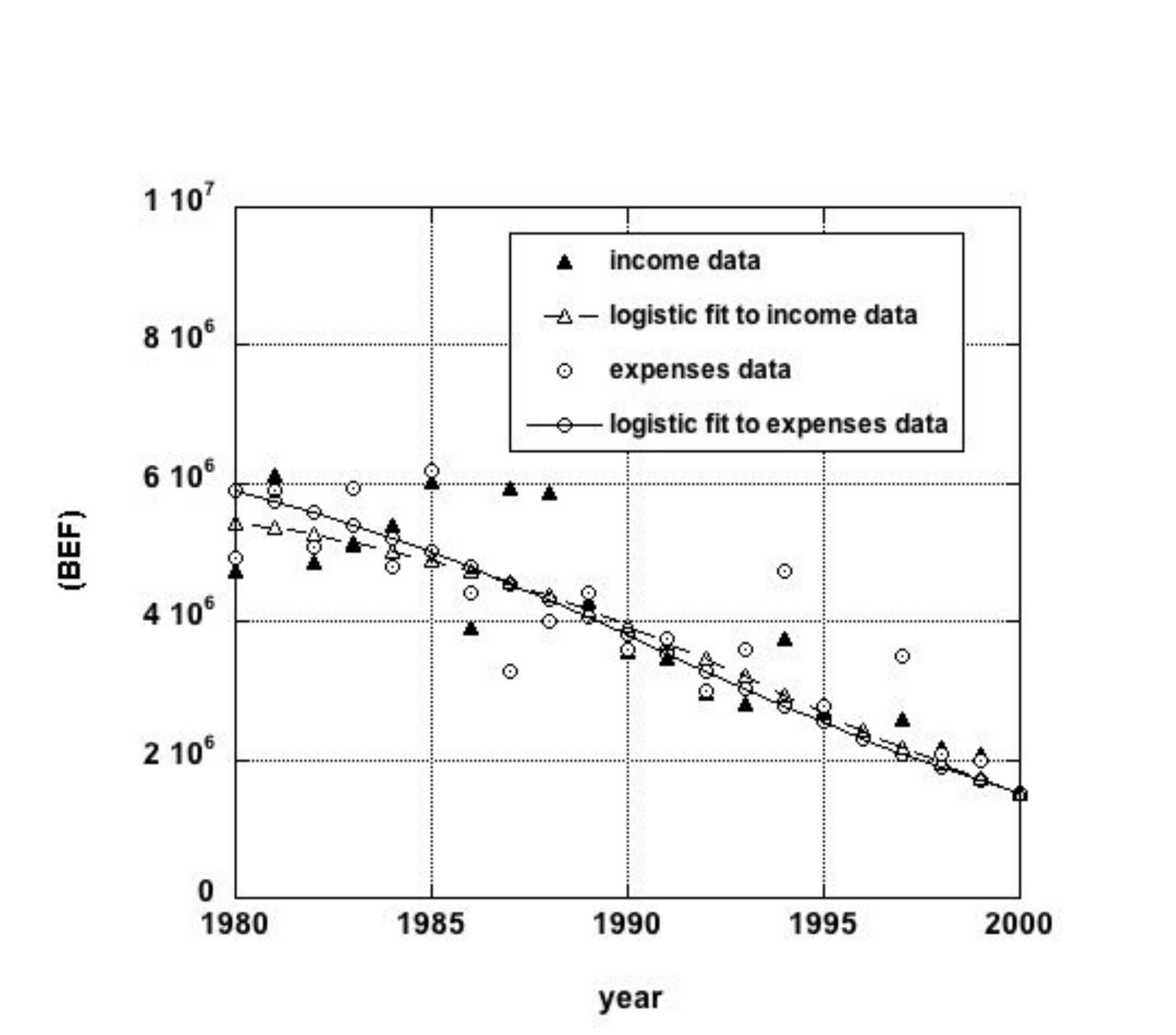}
\caption{Income and expenses data of the Belgium Antoinist Cult Community, from 1980 till 2000,  likely indicating a slow  logistic decay after $ca.$  1980 }
\label{fig:Plot66incexpdecaylogistfitbw}
\end{SCfigure}

Three analytic decay  laws have been tested  to  sketch\footnote{Extreme value   data points have been excluded  from the analysis     indeed for keeping $R^2$ meaningful; compare Fig.\ref{fig:incomexpenses} and Fig.\ref{fig:Plot66incexpdecaylogistfitbw}}
 some relevant relation between the  natural log ($ln$) of the  income and expenses,  after 1975 till (excluding) 2000: (i) a  linear, (ii) a power, and (iii) an exponential law. The following forms, with only two parameters,  were used
 \begin{equation}
\label{decaylog}
  ln(y)= a - b*(year-1975)
\end{equation}
 \begin{equation}
\label{decaypow}
    ln(y)= a*(year)^{-b}
\end{equation}
 \begin{equation}
\label{decayexp}
 ln(y)=a* e^{-(year-1975)/b}
\end{equation}
The best parameter values, with the global $R^2$, are given in Table  2. 
  It appears that the three laws lead to rather similar variations, with $R^2 \sim 0.84$ or $\sim 0.73$ for the income and expenses data respectively. Note that a decaying law like $ ln(y)= a*(year-1975)^{-b}$  does not give anything meaningful.

  \begin{table}\begin{center}
\begin{tabular}{|c|c|c|c|c|c|c|c|c|}
\hline
\multicolumn{4}{|c|}{income }&&\multicolumn{4}{|c|}{expenses  }\\
\hline\hline
&$a$&$b$&$R^2$&&&$a$&$b$&$R^2$\\
\hline
$linear $ & 6.93 &  0.0237   &0.847& & $linear $ &6.87&0.0195&0.729\\
\hline
$power$  &  $2.17 \;10^{24}$ & 7.13   &0.843&&$power$  &$1.68\; 10^{20}$  &5.88& 0.727\\
\hline
$exp.$&6.945&278.5 &0.843  & &$exp.$&6.87&337.8&0.727\\ 
\hline
\end{tabular}
\end{center}
\caption{ Parameter values for  decay laws appropriate to the  yearly income and expenses of the Belgium ACC 
 }\label{incomexpensdecays}
\end{table}

 To remain within a logistic map  philosophy, an {\it ad hoc}  fit has been searched for  on the income  ($y_i$) and expenses ($y_e$) data for the years after 1980 till (excluding) 2000, - turning back the time axis in Verhulst equation. The following results are found 
 \begin{equation}
\label{c9}
y_i \; \simeq  \frac{ 5.9  \; 10^{-6}}{1+e^{0.175*(year-1994)}}
\end{equation}
\begin{equation}
\label{c10}
y_e \; \simeq  \frac{ 7.1 \; 10^{-6}}{1+e^{0.145*(year-1991)} }.
\end{equation}

 
 Yet such expressions and parameter values should be considered as merely indicative ones. They  should be taken with some caution, since the position of the maximum, thus the first decay year of this decay  regime, is rather ill defined in the data, because of  wide fluctuations near this maximum. 
 
 Nevertheless, a definitive conclusion can be reached, after comparing such values with the $r$ value  found, see Table  1,  for the 1968-1980 regime: 
    the growth rate ($\sim0.6$)  is approximately  4 times  larger than the decay rate ($\sim0.15$).  Note that this asymmetry between growth and decay is similarly found in business cycles, see e.g.   \cite{sanglierauslooscycles}. Thus one may suggest here to wonder whether some
 ABM  simulation leading to  such asymmetric features can be imagined. An open question!
 
 One might also wonder whether some revival of  the community, as in the phoenix effects discovered in \cite{phoenix} can appear in the future due to the present political and sociological constraints, and the new forces similar to those known at the beginning of the 20-th century. 

\section{ Conclusions}
 \label{sec:conclusions}

 The complexity of qualitatively studying  agent based groups, like religious communities,  through  their   social and historical  aspects is known, but the quantitative, in particular financial,  aspects are also challenging.  Data is  usually sparse and not necessarily reliable  \cite{579,futures,TAMir}, - except when legal constraints are imposed.   It was of interest  to find a case  with a growth regime $and$ a decay regime, in order to have some insight on the causes of such behaviors for further deep modeling. Facing such a challenge, one goal has been to find a society  evolving on a rather short time span, such that the data be reliable as much as possible.
   
It was of common knowledge that a religious community, the Antoinists, here called ACC, having appeared in the 19-th century, in Belgium, but not so flourishing nowadays,  could present  a bump feature in the number of adepts.  A difficulty stems in the voluntary lack of such an  information in the cult system. However,  the number of temples, since the rising of the first one in  1910 are reliable data, and publication of financial data have been  made mandatory after 1919.
  Their  study has led to interesting features. 
 
  Since the logistic growth function 
   has proven useful in modeling a wide variety of phenomena of growing systems, it has been used as the analytic equivalent of an ABM.
However, complex social  systems rarely follow a single S-shaped trajectory. They often present  a simple extremum or (irregular) oscillations,   whence implying the need  to go beyond  simple Verhulst-like models     \cite{577PhA391.12.3190Ð97auslantoin,AusloosGV}.  Here, it
 has readily appeared  that two regimes must be investigated for the temple number evolution: (I) one between 1010 and 1919,  for 16 temples; (II) another between 1923 and 1935, for 11 temples. Three (asymmetric) regimes are found for the financial evolution.
 
 The "model" indicates that such communities are markedly influence by external considerations ("external  fields"),
beside their intrinsic "religious" goals. Practically, in the present case, as illustrated, the crash of 1929 induces a drop in income, but the second world war
increases the community strength. The golden sixties "reduce" the income: the
adepts wealth is being increase, but they reduce their offering, becoming in some
sense more egoistical. Therefore, one can deduce that there are two different
causes for the drop in income: either a lack of money of the adepts, or in contrast, paradoxically, "too much" wealth. Similarly, the increase in cult income,
at its legal beginning, may result from the thanking for healing the suffering, -
but also occur due to the income explosion till 1985. The variation in expenses
are immediately related with such income considerations.
Therefore such an ABM,   apparently leading to a   "universal"-like  interpretation, contains ingredients,
with non-universal "amplitudes", 
but  is expected to be applicable to other societies,
 
 From a  practical point of view, it has to be emphasized that
the ACC  was appealing because of the suffering of people, working
under very hard conditions in the Li\`ege,  BE, area.  The catholic social system was lacking convincing impact. Local people were searching within proto-science  appeal for mind and soul healing through connections with spiritualistic phenomena. Thus, within a pure altruism, P\`ere Antoine started to
preach and to give psychological remedies, i.e.,  "first principles", for accepting one's life, sometimes  "demonstrating"  his  ''body healing powers''. The initial seed of the ACC grew within a rather weak competitive framework, due to  P\`ere Antoine's charisma, simplicity, and affection.  A follow-up by local people resulted, not far from  recalling what  happened  a long time ago, in what are now more established religions.  However,  his charismatic  leadership was most likely  lost,  at P\`ere Antoine's death.  Moreover, there was not much serious attack, nor martyrs appearing, which are two aspects for an increase in community strength and expansion  \cite{roachbook}.  No cult of  saints  was  established, on one hand, and on the other hand, social and economic conditions largely improved, reducing the need for intra-community self-support.  Therefore some size leveling off had to be expected; as well seen in the evolution  of the  number of temples.

A marked
decay, in adepts, is known to  occur  since the end of the 20-th century, inciting one to conclude to a
doomed situation, according to the theory of Stark  \cite{Stark96b} - in contrast to, e;g.,  the Jehovah's 
Witnesses \cite{StarkIannaccone97jehovah}. Indeed, ideologies (whether religious or secular) seem to lack
coherence and potency unless they are developed and promulgated by vigorous
formal organizations and social movements. 
 One crucial aspect of the ACC  concerns
its survival under much improved economic, social, and health conditions of
workers to whom the P\`ere Antoine "philosophy" appealed. 
 
 Finally, one might nevertheless wonder  about the future of the community, i.e. whether some revival of the community,  through  phoenix-like effects    \cite{phoenix} can appear in the future within present political, economic, and sociological conditions, upon the constraints of new forces  on workers  having hard job conditions,   though not immediately similar  to  those known at the beginning of the 20-th century.

  Thus, {\it as should most likely be  really expected in fact}, it is concluded that social phenomena are very complex processes offering much challenge  for  quantitative mathematical  modeling. Nevertheless the above findings 
   lead  to simple empirical laws.  Even though the ACC is surely a marginal religious case, the merging of qualitative and quantitative considerations as done here above might appeal to finding other cases of interest among sects having reliable data and  do suggest generalizations.

  \bigskip
{\bf Acknowledgements} 
Comments on preliminary versions by J. Hayward, A. P\c{e}kalski, and F. Schweitzer have surely improved the present paper. Infinite thanks go to Fr\`ere  W. Dessers  and Soeur S.  Taxquet,  
presently President and Secretary of the Administration Board of the Antoinist Cult in Belgium,  respectively, for their kindness, patience,  availability, when I asked for data and historical points, and for  trusting me, -  when allowing me to remove archives from the library for scanning outside the office.

\begin{center}
  {\bf Appendix A: Historical Perspective}\label{sec:App_history}
\end{center}

Louis Antoine was born in  1846 near Li\`ege, Belgium, the youngest son of a 8 children family, baptised as catholics; his father was a coal miner. He followed the usual catholic education of the time, read much,  became a coal miner and later on a metallurgist worker. After killing inadvertently a friendly soldier during his military service, he was punished, but used that accident as a basis for thinking about "good and bad"; he volunteered to work abroad for the Cockerill factory; got married, and had a son. At 42, he got more sedentary, continued to think much about religion, mankind role and life values, He read Kardec's Livre des Esprits (1857) and was enthused. He wishes to become medium becomes the founder of a spiritism group "Les Vignerons du Seigneur", but his son dies at 20 in 1893. He begins  to discover that he can be ''healing'', writes a small book about healing, gathers friends and starts becoming a cult leader, and is known as a healer. But never asks any money for his miracles. Nevertheless he gets into trouble with justice, and medical doctors. Somewhat to avoid such problems, he reminds the patients that faith is the key; he transfers his good health into others by faith. Still gets problems with the law: he gets 1200 people per day at home for doing some so called ''operation''. Later on, considering that to heal a body is not enough, he turns toward more moral value rebirth. His predicator role increases, in the Jemeppe ''temple'', and in 1906 starts publishing  notes taken by some scribe, outlining his doctrine. The concept of disease is denied, just as is that of death, and there is belief in the reincarnation: it is intelligence which creates suffering;  only "faith in"    removes it, and not the intervention of health professionals. He becomes  the ''Father'', (le P\`ere) and establishes some clothing rules (a long black dress for ''desservants''). At Easter 1910, he is considered as a ''prophet'', and on Aug. 15, 1910 sanctifying the first temple. He will expand and clarify the doctrine ({\it Evil does not exist}), will raise more temples till his death in 1912. His wife, Catherine, who could not read,  maintained and pursued the cult activities amongst schisms and heresies, till her death in Nov. 1940.
 
 To put several references in evidence,   let a few be quoted: 
\cite{Debouxhtaybook,Dericquebourg,AntoineL,Vivier}  
where much can be found on the doctrine, rules  (dressing),  symbols (the tree), prophecies,   testimonies,  social roles and values, spiritual and philosophical (reincarnation)  contexts, also with  some historical and sociological perspectives; see also  $www.culteantoiniste.com$.

In essence, let it be emphasized that there is no search for increasing the financial  wealth of the hierarchy. There is some 8-year rotation in duties. It is known that a check, used as an  offering, is not cashed in a bank because it has a personal connotation. Only anonymous offerings  are accepted; there is no proselytism and one does not ask for money from followers. The ''desservants'' are not paid.  There is neither exclusivity on religious adherence nor  does one provide any prescription on social and political issues. The main goal is to worship and to heal, in line somewhat with Mary Baker Eddy's ideas \cite{MaryBakerEddy}.

In my opinion, the Antoinism, or the Antoinist Cult,   cannot be considered as a sect, - though it was so in France ; see Journal Officiel,  Commission d'enqu\^etes parlementaires sur les sectes en France, Rapport 2468, Dec. 1995   \cite{JOFrance}. Is it  a religion, a church? Maybe, but there is no  strong structural hierarchy and clerical body. It is surely, from a catholic	 christian  point of view, a heresy. Is it a cult? in the sense of Nelson \cite{Nelson}, most likely yes;  it should therefore disappear if there is no one to pick up the ideas and turned them toward some money getting  religious scam.





\begin{SCfigure}[\sidecaptionrelwidth][h!]
\centering
  \includegraphics[height=5cm,width=8cm]{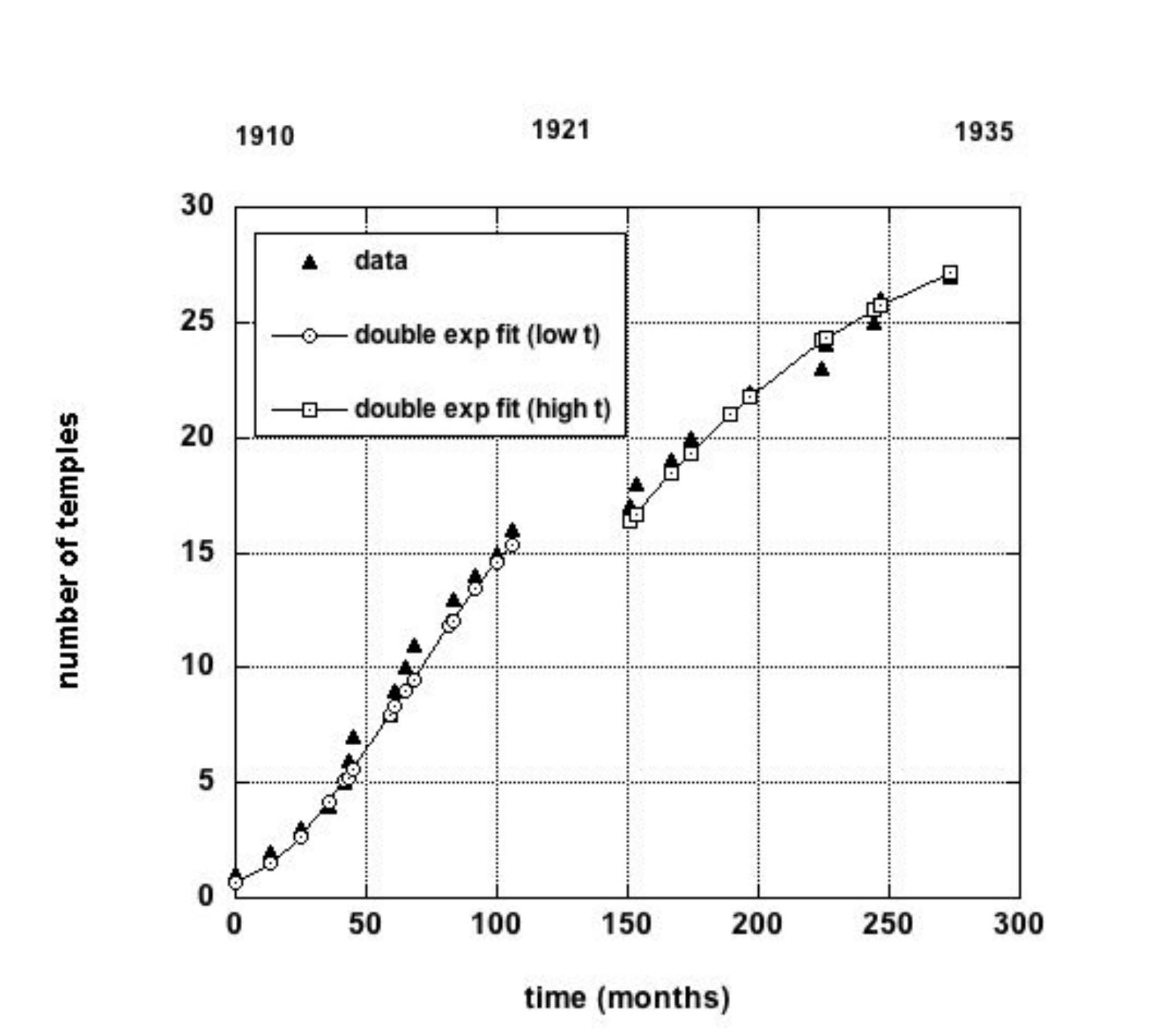}
\caption{
  Gompertz double exponential law  fit of the number of consecrated ACC temples in Belgium as a function of the number of months (cumulated), in  low and high $t$ regimes  since the consecration of the first temple in 1913 }
\label{fig:PlotG31AKLFtemplefitsbw}
\end{SCfigure}
 
 \begin{center}
  {\bf Appendix B: Gompertz Double Exponential  Law}\label{ANTEGDEL}
  \end{center}
  
  Similarly to the  analysis reported in the main text, a  Gompertz double exponential law fit can be searched for the number of temples as a function of the number of months (cumulated again).
 
 The best fit  behavior  to such a Gompertz double exponential law has been searched through a log-log plot method, imposing the amplitude to be an integer. It has occurred  after many simulations that  two distinct regimes must be considered, exactly as in the analysis along the  Verhulst  approach: one at {\it low} time, i.e. during the initial growth of the ACC, and another at later ($high$)  time, with a 4 year gap, between 1919 and 1923. One obtains respectively : 
\begin{equation}
\label{Glowt}
   c(m)=23\;e^{-e^{-(m-62)/48.5}}
\end{equation}
 \begin{equation}
\label{Ghight}
  c(m)=31\;e^{-e^{-(m-116)/77.5}}
\end{equation}
 as shown on Fig. \ref{fig:PlotG31AKLFtemplefitsbw}. Note that the upper  ($absolute$) values of the possible number of temples to be expected slightly differ in the Verhulst and Gompertz approaches, -  though  in an  opposite $relative$ value for the low and high $t$ regimes.  The rates found in Eqs. (4)-(5) and here above, in Eqs. (12)-(13),  are comparable: 0.034 $\sim $  (1/48.5 $\simeq$) 0.021, and 0.0195 $\sim$ (1/77.5 $\simeq$) 0.0129 respectively, depending on the regime.

\begin{SCfigure}[\sidecaptionrelwidth][ht!]
\centering
  \includegraphics[height=7. cm,width=8.5cm]{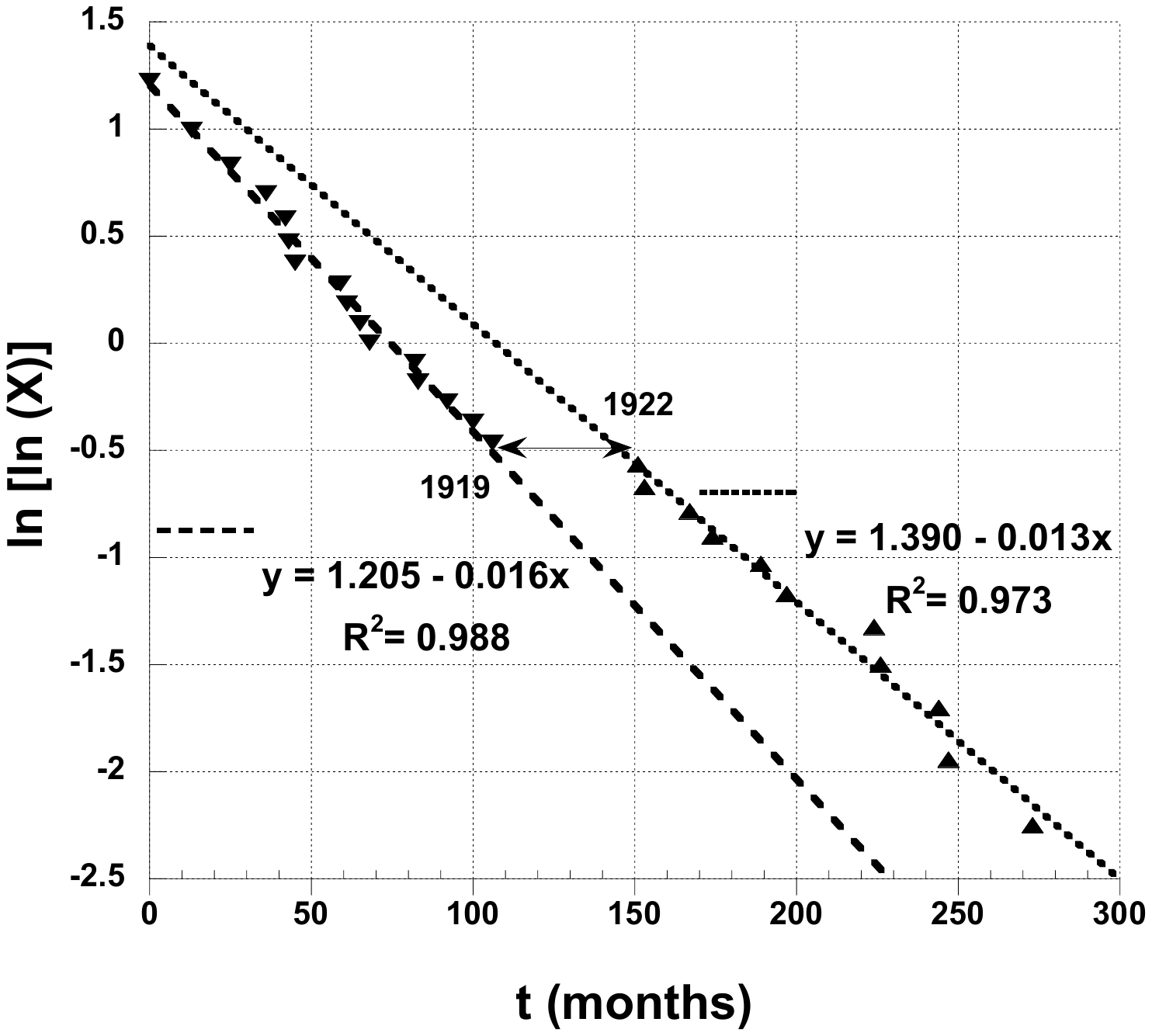}
\caption{  Gompertz plot: Log(Log) variation of the (relative to some maximum possible) number of ACC temples ($X$) raised  in Belgium as a function of the number of months, starting from the raise of the first temple 
 indicating a "social force effect" influencing a variation in growth rates
  }
\label{fig:Plot59templeGforcellt}
\end{SCfigure}

\begin{center}
  {\bf Appendix C: On Social  Forces}\label{sec:social_history}
\end{center}
 
It is hereby  emphasized that social forces can be introduced at least in two different ways in an ABM, based on Vehulst and/or Gompertz analytic evolutions.
 Indeed, a different  adaptation of the ideas in  \cite{PNAS75.78.4633-7-Montroll-socialforces}  on the evolution of competing entities, economic or sociologic ones,  occurs if,  instead of  Eq.(\ref{Montrolleq}),  one writes
\begin{equation}
\label{Montrolleq2}
\frac {d\;ln(X)}{dt }= k(1- X) +  ( \alpha /\;\theta) (1- X) \;.[H(t-\tau)- H(t-(\tau+\theta)]
\end{equation}
where $H(t)$ is the Heaviside function. In other words, one is (mathematically)  letting Montroll's $\alpha$ to be $X$-dependent over the time interval [$\tau; \tau+\theta$].  However, the emphasis  differs much from \cite{PNAS75.78.4633-7-Montroll-socialforces}: i.e.,  rather than modifying the (Malhtus) $X$ term, one adapts the (Verhulst) $(1-X)$ term, to (economic or social) constraints.

Mathematically, on the same footing as  Eq.(\ref{Montrolleq2}), one can write 
\begin{equation}
\label{Montrolleq2V}
\frac {dX}{dt }= kX (1- X) +  ( \alpha /\;\theta) X (1- X) \;.[H(t-\tau)- H(t-(\tau+\theta)].
\end{equation}

Readily, the rate before the pulse  is $k$ but is $k+\alpha/\theta$   after the "pulse" application.  This "second rate" depends on the pulse strength and some  time duration $\theta$.

Within the Gompertz framework, starting from the differential equation
\begin{equation} \label {Gompertz1dif}
\frac{dy}{dt}= r \; y \;  log\left[  \frac{k}{y} \right]\;\\,
\end{equation}
it is "sufficient" to replace $(1-X)$ by $\sim -ln(k/y)$; compare Eq. (\ref{Montrolleq}) and Eq.(\ref {Gompertz1dif}).
A double log ($X$) plot as a function of time  is shown in    Fig. \ref{fig:Plot59templeGforcellt}, i.e. an appropriate replot of Fig. \ref{fig:Plot53templeforceAB}, - the best fit  equations being  written in the figure. The fit is very precise. It is emphasized that the fit  lines are $not$ parallel anymore. From these, one can deduce $\theta_1= 34$ (months)  and $\theta_2= 135$ (months), i.e. $\sim$ 3 and 11 years respectively.

\end{document}